\documentclass[aps,prl,twocolumn,showpacs,superscriptaddress]{revtex4-1}  

\usepackage{graphicx}  
\usepackage{dcolumn}   
\usepackage{bm}        
\usepackage{amssymb}   
\usepackage{mathtools}
\usepackage{etoolbox}
\usepackage{braket}
\usepackage{subcaption}
\appto\normalsize{\belowdisplayshortskip=\belowdisplayskip}
\appto\small{\belowdisplayshortskip=\belowdisplayskip}
\appto\footnotesize{\belowdisplayshortskip=\belowdisplayskip}
\usepackage[utf8]{inputenc}
\usepackage{color}
\usepackage{soul}
\DeclareMathSizes{5}{13}{7}{3}

\begin{document}

\title{Measurement induced phase transitions in quantum
raise and peel models}
\author{Eliot Heinrich}
\address{Department of Physics, Boston College, 140 Commonwealth Avenue, Chestnut Hill, Massachusetts 02467, USA}
\author{Xiao Chen}
\address{Department of Physics, Boston College, 140 Commonwealth Avenue, Chestnut Hill, Massachusetts 02467, USA}

\date{\today}

\begin{abstract}
We present a quantum circuit model which emulates the interface growth of the classical raise-and-peel model. Our model consists of Clifford unitary gates interspersed with projective measurements, applied according to prescribed constrained update rules. We numerically find via large-scale simulations that, depending on the constrained update rules, the system may undergo several measurement-induced entanglement transitions, including continuous transitions as well as a second phase transition which we interpret using a biased random walk picture. We present a quantum circuit model which emulates the interface growth of the classical raise-and-peel model. Our model consists of Clifford unitary gates interspersed with projective measurements, applied according to prescribed constrained update rules. Through large-scale numerical simulations, we find that the system, depending on the constrained update rules, may undergo several measurement-induced entanglement transitions. These include continuous transitions observed in  previously studied hybrid circuits and a phase transition, which we interpret using a biased random walk picture.
\end{abstract}

\maketitle

\section{Introduction}
The past few years has witnessed the rapid development of quantum information dynamics, including the information scrambling and the emergent light cone behavior~\cite{shenker2015stringy,Maldacena_2016,Chen_2019,zhou2020,Xu_2019,Nahum_2018,von_Keyserlingk_2018,Chen_Lucas_2019,Tran_2019,Lashkari_2013,Nahum_2017}. It is established that for many-qubit systems undergoing local unitary evolution, quantum systems tend towards highly entangled ``volume-law" state, in which the entanglement entropy of a subsystem scales proportional to the volume of the subsystem. 

Volume-law entanglement entropy scaling can be mitigated by interspersing the unitary evolution with projective measurements. Interestingly, when the measurement rate is low, a highly entangled volume-law phase persists. However, increasing the measurement rate induces a phase transition to a disentangled ``area-law" phase, where the entanglement entropy of a subsystem scales proportionally to the boundary area. This phenomenon, known as the measurement-induced phase transition (MIPT), has been extensively studied, with the construction of numerous one-dimensional circuits to elucidate its underlying physics~\cite{skinner2019, li2018, chan2019, koh2023, zabalo2022, iaconis2020,choi2020,Gullans_2020,li2019}. An intuitive conceptualization of this phase transition is offered through the surface growth picture, where the entanglement entropy, represented as a ``height", undergoes local increases via the application of a local unitary gate, while single-qubit measurements, on average, locally suppress the height~\cite{Nahum_2017,li2019,Nahum_2018,morral2023}. The competition between the two leads to the emergence of MIPT.

An example of a well studied classical surface growth model is the raise-and-peel model (RPM)~\cite{degier2004, jara2018}. In the RPM, a substrate on a 1$d$ lattice is updated by either locally depositing to the substrate or nonlocally ``peeling" a layer, depending on the local configuration of the substrate. The RPM thereby has a {\it feedback} mechanism, wherein the configuration of the substrate affects dynamic updates applied in the immediate future. Analogously to the MIPT, this competition between adsorption and desorption leads to phase transition between a ``flat" phase in which the average substrate height remains finite, and an ``adsorbing" phase, in which the average substrate height is proportional to the system size. 

The RPM additionally has an intermediate phase between the flat and adsorbing phases in which the substrate height scales algebraically with the system size, though slower than linear with the system size~\cite{degier2004}. Both the adsorbing and intermediate phases are characterized by long power-law tails in the sizes of avalanches, indicating that this model is an example of self-organized criticality~\cite{degier2004, alcaraz2006}.

Building on the aforementioned research, this paper constructs a quantum entanglement entropy analogue to the RPM, replicating its feedback mechanism through non-unitary quantum dynamics. We focus on Clifford circuits, so that the entanglement entropy takes integer values in the logarithm base 2 and we are able to simulate the system for relatively large numbers of qubits. Application of local unitary gates emulates a strictly local surface deposition event, or a ``raise" in the language of the RPM, while a projective measurement emulates a nonlocal ``peel" update. We define a group of constrained update rules which determine when a unitary/measurement gate should be applied, thereby creating a feedback mechanism between the entanglement substrate and subsequent update rules.

We find that the details of the update rules plays a very important role in the resulting phase diagram of the system. We identify three nontrivial phase diagrams, shown in Fig. \ref{fig:qrpm_model}(c), where two include continuous phase transitions. We also observe new phase transitions from the area-law and nonthermal volume-law phases into an almost maximally entangled phase. In this phase, the entanglement entropy grows linearly with time, accompanied by diffusive fluctuations. These dynamics can be effectively described in terms of a 1d biased random walk. 

\section{Model description}

\begin{figure*}
  \centering
  \includegraphics[width=\linewidth]{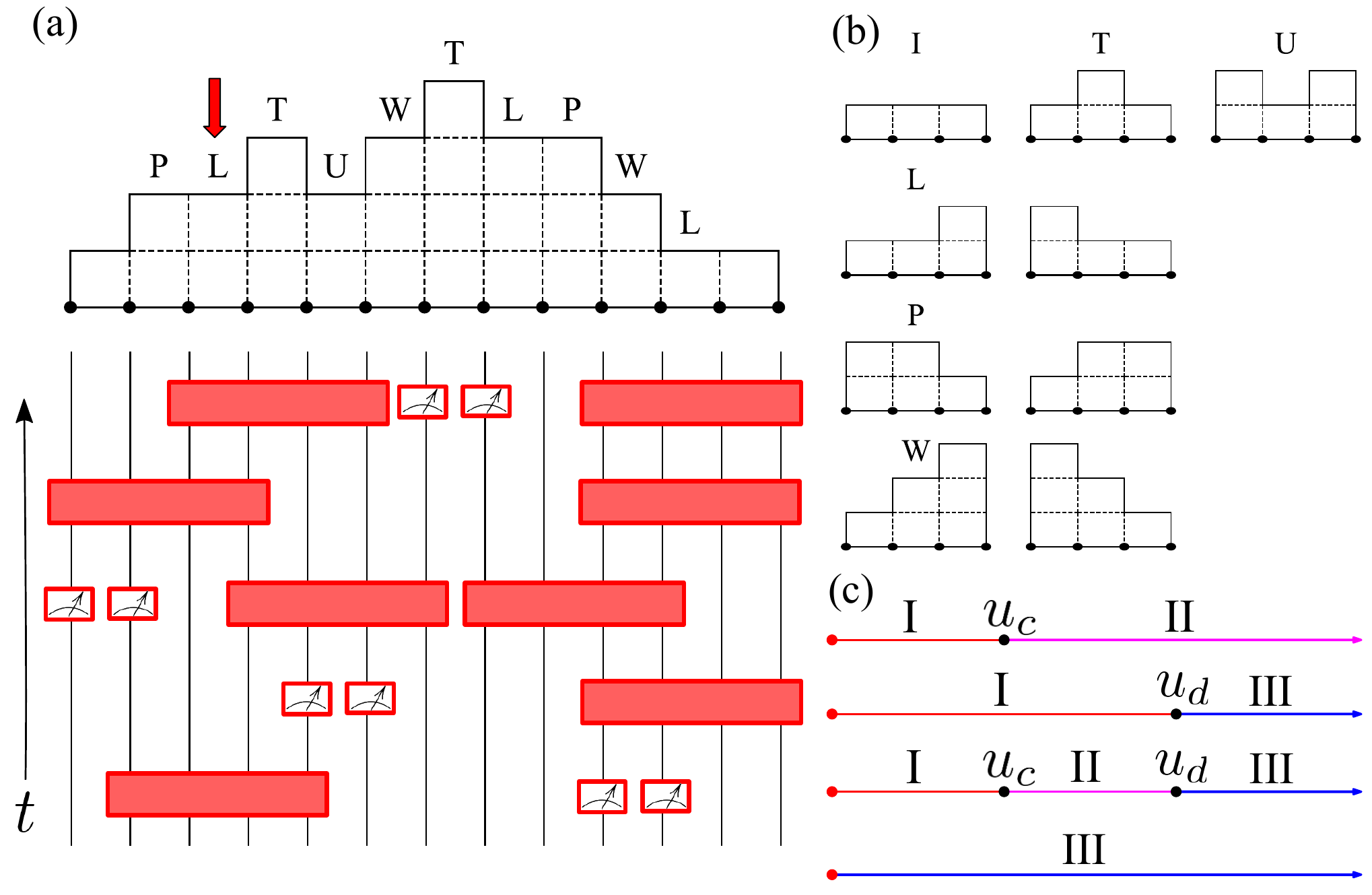}
  \caption{(a) The entropy substrate defined in the text. The top half of the figure show the entropy substrate $h_i$, with the blocks labelled. The qubits in the underlying quantum state are indicated with black circles. At each timestep, a site on the substrate is randomly selected; an example is indicated by the red arrow. Depending on the block observed, in this case an L-block, an update may be applied. If the observed block indicates a unitary operation, a four-qubit Clifford gate is applied to the four corresponding qubits with probability $p_u$. Otherwise, measurements in the $Z$ basis are applied to the two corresponding qubits with probability $p_m$. In the bottom half of the figure, the corresponding quantum circuit evolution is displayed with commuting operations drawn as occurring at the same timestep. (b) The possible blocks specifying the local structure of the entropy substrate are labelled I, T, U, L, P, and W. Note that the L, P, and W blocks each correspond to two distinct blocks which are equivalent under reflection, preserving directional symmetric in the update rules. (c) The possible phase diagrams for the model, depending on the update rules. The model parameter $u$ increases from left to right. At $u = 0$, the entanglement entropy follows an area law regardless of the feedback mode, as no unitary operations occur. The area law (I) is indicated in red. In the first row, as $u$ is increased, the system undergoes a continuous phase transition at $u = u_{c}$ into a strongly fluctuating nonthermal volume law phase (II); this volume law is indicated in magenta. In the second row, the system instead undergoes a phase transition at $u_d$ to a phase where the entanglement grows linearly with time, accompanied by diffusive fluctuations. After long-time evolution, it freezes into an almost maximally entangled volume-law state with negligible fluctuations. This phase (III) is indicated by blue. In the third row, the system undergoes both the continuous and phase transition characterized by diffusive fluctuations, whereby II becomes an intermediate phase. Finally, in the last row, the system undergoes no phase transition, staying in the frozen volume law III for all $u > 0$. In the main text, we focus on representative models that exhibit phase transitions, and show results for other models in Appendix A.}
  \label{fig:qrpm_model}
\end{figure*}

\begin{figure*}
    \centering
    \includegraphics[width=\textwidth]{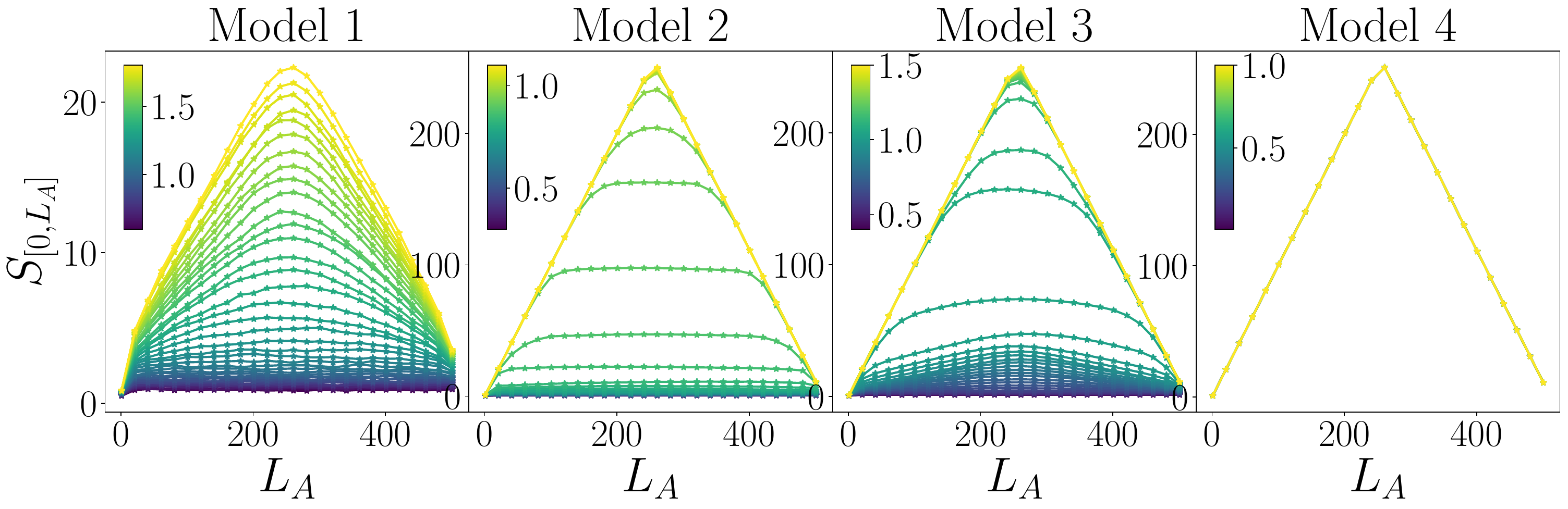}
    \caption{The entropy substrate for an update rule configuration mentioned in the text. For each model, a system of $L = 512$ qubits is evolved starting from an initially unentangled state (except for model 4, which starts from a fully entangled state) for $3000L$ steps to reach static equilibrium, followed by $2000L$ steps during which the entropy is sampled. We again average over 20 circuit realizations. The parameter $u$ is selected uniformly on an interval $[u_{\text{min}}, u_{\text{max}}]$, with the purple lines corresponding to $u = u_{\text{min}}$ and the yellow lines corresponding to $u = u_{\text{max}}$. For model 2, the curves lying between the area-law and volume-law scaling are near the transition point of the model, as described in the text, have very long equilibration time ($t \sim L^2$). As a result of limited computational resources, some of these states very close to the critical point have not reached their final, maximally entangled state. A finite-size scaling analysis is presented in Fig.~\ref{fig:rep_crossings}. Note that fluctuations are absent in the steady state of the maximally-entangled volume law phase observed in models 2, 3, and 4, as indicated by the error bars. }
    \label{fig:rep_modes}
\end{figure*}

Consider a bipartition of a quantum state on a one-dimensional lattice of $L$ qubits subject to open boundary conditions into subsystems $A$ and $B$. The Renyi entanglement entropy of subsystem $A$ is computed as 
\begin{align}
    S_A^{(\alpha)} = \frac{1}{1-\alpha}\log_2 \text{Tr}\left( \rho_A^\alpha \right), \quad \rho_A = \text{Tr}_B \left( \rho \right)
\end{align}
 where $\rho = \ket{\psi}\bra{\psi}$ is the density matrix of the quantum state $\ket{\psi}$. For the remainder of this work, we will focus on stabilizer pure states~\cite{aaronson2004}, for which the Renyi entanglement entropy takes integer value and is independent of Renyi index $\alpha$. We will drop the $\alpha$ from the notation. We will further restrict our interest to bipartitions in which $A$ is a connected interval of qubits denoted $A = [a, b]$, where $a$ ($b$) is the leftmost (rightmost) qubit of $A$. The entanglement entropy of this interval is denoted $S_{[a, b]}$. We define the entropy ``substrate" height as $h_i = S_{[0,i]}$, where $i$ is a site on the lattice dual to the qubit lattice, as shown in Fig. \ref{fig:qrpm_model}(a). Note that $h_i$ automatically obeys the restricted solid-on-solid constraints typically enforced in interface growth models~\cite{kim1989}, as the entanglement entropy can change by at most 1 when a single qubit is added or removed to the interval, and $h_0 = h_L = 0$ by definition. As the states of interest are stabilizer states, the entanglement entropy profile is further restricted to change by $h_{i+1} - h_i \in \{-1, 0, 1\}$. 

Acting on the entanglement substrate, we define update rules in the spirit of the RPM. For each discrete time step, a random site $j \in \{1, 2, ..., L-2\}$ is chosen. Based on the configuration of $h_{j-1}$, $h_j$, and $h_{j+1}$, either a four-qubit random Clifford gate, sampled using Ref.~\cite{berg2020}, or two single-qubit projective measurements may be applied. We apply four- and two-qubit operations respectively to respect the fact that in the substrate picture, a region of three sites corresponds to four qubits, and the operations must act on an even number of qubits to avoid biasing operations in one direction or another. The four-qubit Clifford gates generically increase the entanglement in the region they act on, emulating a ``raise" update in the RPM, whereas the projective measurements are able to affect the substrate nonlocally, emulating the ``peel" update. As an extreme example, measuring any qubit in a GHZ state affects the entanglement substrate at all points in space.

The local substrate configuration can be classified according to six possible (up to reflection) blocks, enumerated in Fig.~\ref{fig:qrpm_model}(b). A set of update rules are defined by a set of blocks for which a unitary gate is applied with probability $p_u$. For the remaining blocks, projective measurements in the $Z$ basis are applied with probability $p_m$. At each timestep, a site is randomly chosen to be updated according to the feedback mode. The mode as well as the parameter $u = p_u/p_m$ determine the equilibrium behavior of the substrate. We remark that, as the update rules are determined by the local entanglement substrate, and the action of each update rule on the entanglement substrate is not known, the full quantum state must be known at each timestep to re-construct the substrate.

We find that all possible update rule configurations can be classified into four distinct phase diagrams, pictured in Fig.~\ref{fig:qrpm_model}(c). In the text, we will focus on four concrete sets of update rules which are representative of the possible phase diagrams, and show some results for all possible update rules in Appendix A.

\section{Numerical results}

\begin{figure}[b!]
    \centering
    \includegraphics[width=0.5\textwidth]{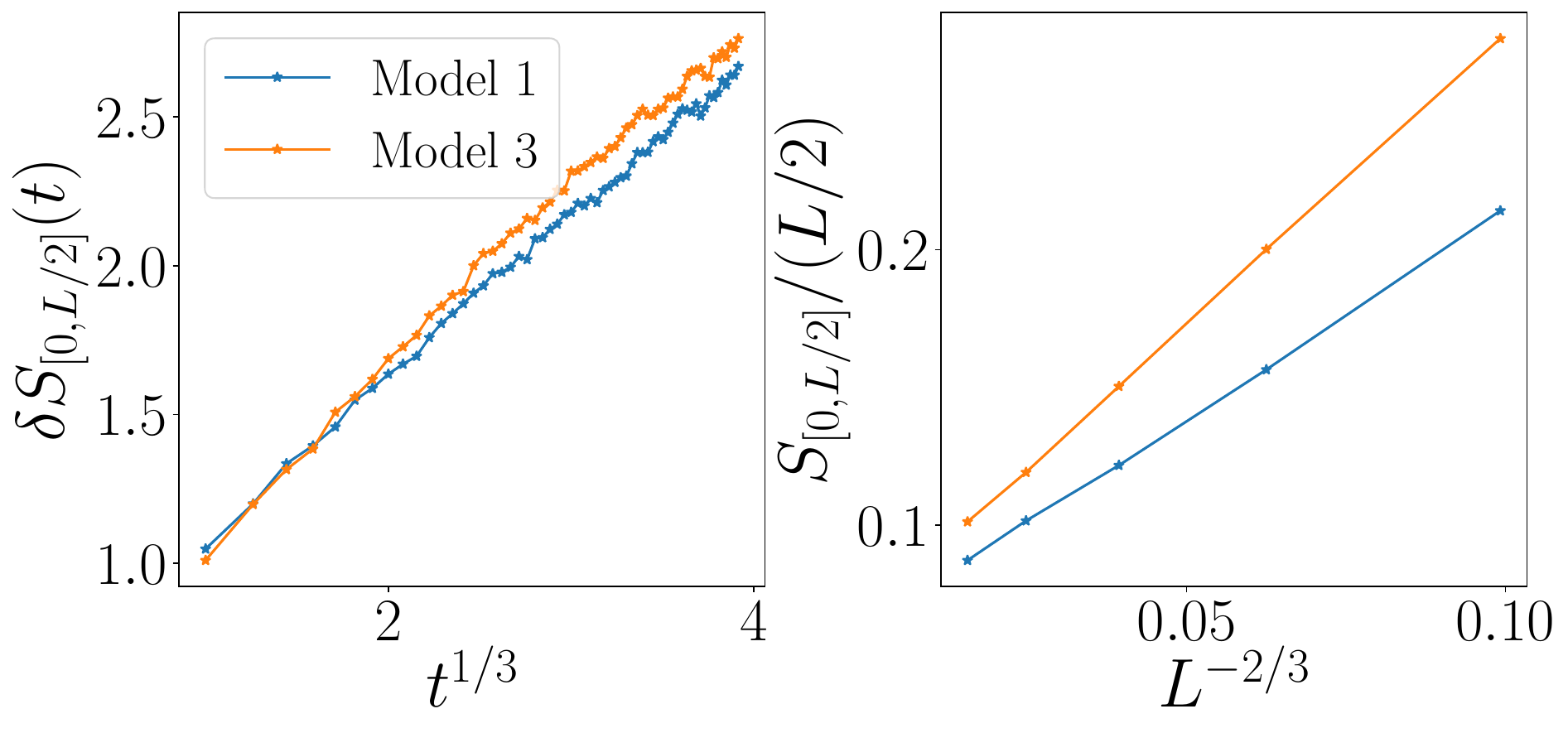}
    \caption{We show the entanglement entropy density of models 1 and 3 against $L^{-2/3}$ for {$u = 1.8$} and {$u = 0.88$}, respectively, which lies within the volume-law of model 1 and the intermediate volume-law of mode 3. The relationship is very close to linear, indicating that the volume-law phase of model 1 and the intermediate phase of model 3 both have volume-law entanglement scaling with KPZ fluctuations. }
    \label{fig:kpz}
\end{figure}

\begin{figure*}
    \centering
    \includegraphics[width=\textwidth]{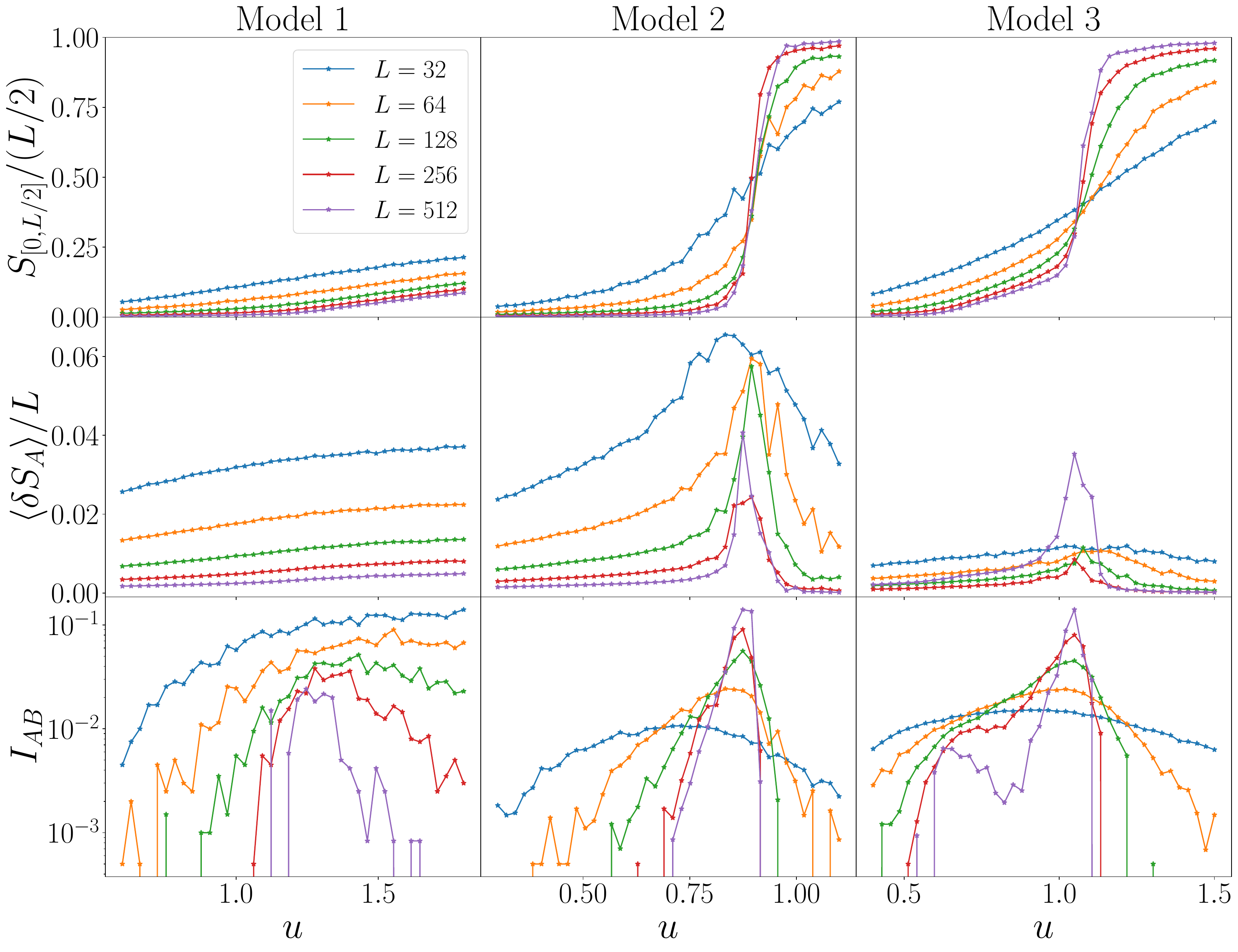}
    \caption{Top: For varying $u$ and $L$, the half-system entropy is plotted. For model 1, the half-system entanglement changes continuously across the phase boundary around $u \approx 1.2$. Models 2 and 3 exhibit discontinuous phase transitions in the half-system entanglement, indicated by the transition becoming increasingly sharp as the system size is increased. Middle: The entanglement substrate fluctuations averaged over the entire profile. Entering the ``frozen" volume law phase in models 2 and 3, described in the text, the fluctuations become strong at the transition point before vanishing in the volume-law phase. Bottom: The mutual information for various system sizes is plotted. The subregions $A$ ($B$) are chosen to be the $L/8$ leftmost (rightmost) qubits. Peaks which sharpen with increasing system size are visible for all three models, indicating the presence of entanglement transitions. An additional second peak is observed for model 3 when $L$ is sufficiently large ($L = 512$), indicating that the system undergoes a second transition as $u$ is increased. We note that the magnitude of the peak near $u = 0.6$ is considerably smaller than the peak at $u = 1.0$, and is only visible by eye on a logarithmic scale.}
    \label{fig:rep_crossings}
\end{figure*}

\begin{figure}[t!]
    \centering
    \includegraphics[width=0.5\textwidth]{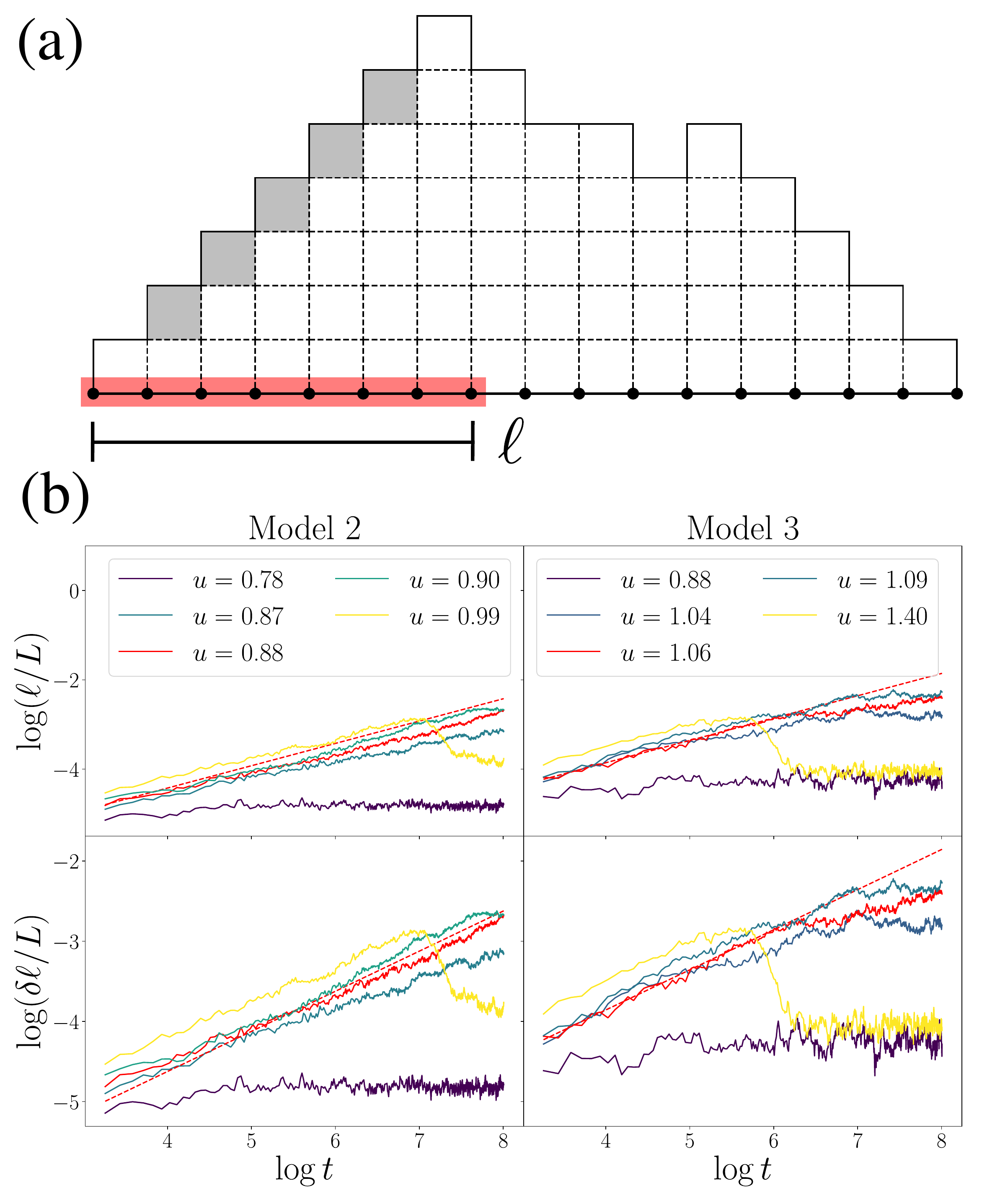}
    \caption{ (a) We show an example of a substrate profile which, if the update rule under consideration applies four-qubit unitaries to the W block, contains a staircase region which is protected from updates except for at the boundaries. This region has length $\ell$. The red box represents the region on which a four-qubit random Clifford gate is applied in our quantum hybrid model and the shaded blocks represent the area of the entanglement substrate which is protected by the RSOS contraints. (b) We show the size of the staircase region and the corresponding fluctuations as a function of time $t$. The system size is fixed to $L = 256$, and results are averaged over 200 circuit realizations. At $u_d$, the length and magnitude of fluctuations of the protected region grows diffusively; the dotted red line has slope of 1/2, and serves as a guide to the eye. Above $u_d$, the length grows with a power greater than 1/2, but continues to fluctuate diffusively, suggesting an effective drift-diffusion description of the size of the protected region. }
    \label{fig:frozen_mode}
\end{figure}

\begin{figure*}
    \centering
    \includegraphics[width=\textwidth]{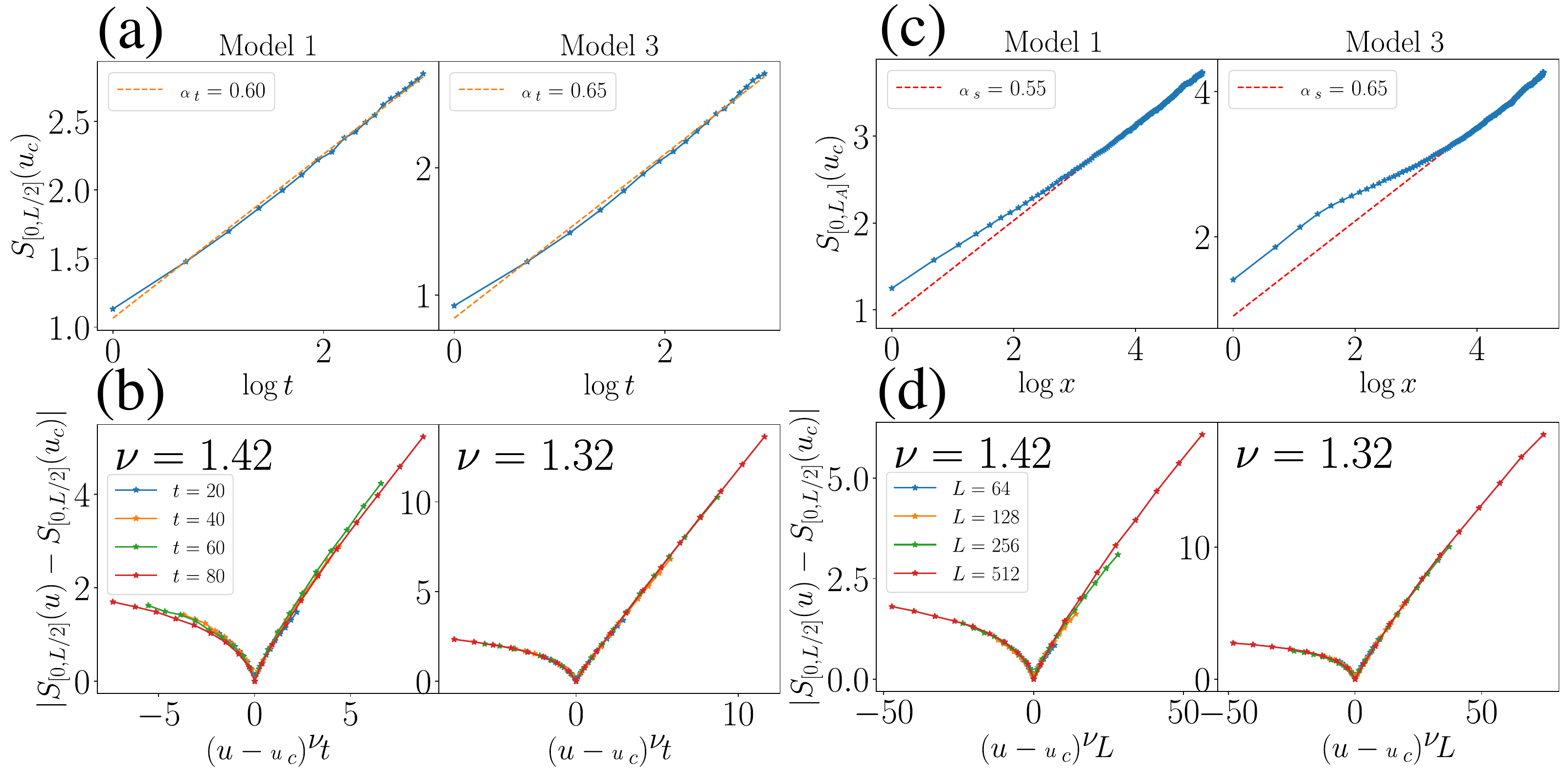}
    \caption{(a) The half-system entanglement entropy before saturation is plotted against $\log t$ for models 1 and 3 at the critical points $u_c = 1.18 \pm 0.02$ and $u_c = 0.62 \pm 0.01$, respectively. The system size is fixed at $L = 512$ and results are averaged over $1.5 \times 10^4$ circuit realizations. (b) The temporal data collapse used to compute the critical exponent $\nu$ for models 1 and 3. We use the same sampling procedure as in (a). For model 3, only the region $u < u_d$ below the frozen volume law are considered in the data collapse. (c) The saturated entanglement entropy is plotted against $\log x$ at the critical points for models 1 and 3. The data is generated by performing $6000L$ timesteps followed by averaging over $3000L$ timesteps and 60 circuit realizations. The first 20 points are excluded from the fit to avoid the boundary behavior stemming from the onset of the protected regions described in the text. (d) The spatial data collapse for models 1 and 3. We use the same sampling procedure as in (c). We observe that model 3 deviates slightly from linear behavior. }
    \label{fig:critical_modes}
\end{figure*}

\begin{figure*}
  \centering
  \includegraphics[width=\linewidth]{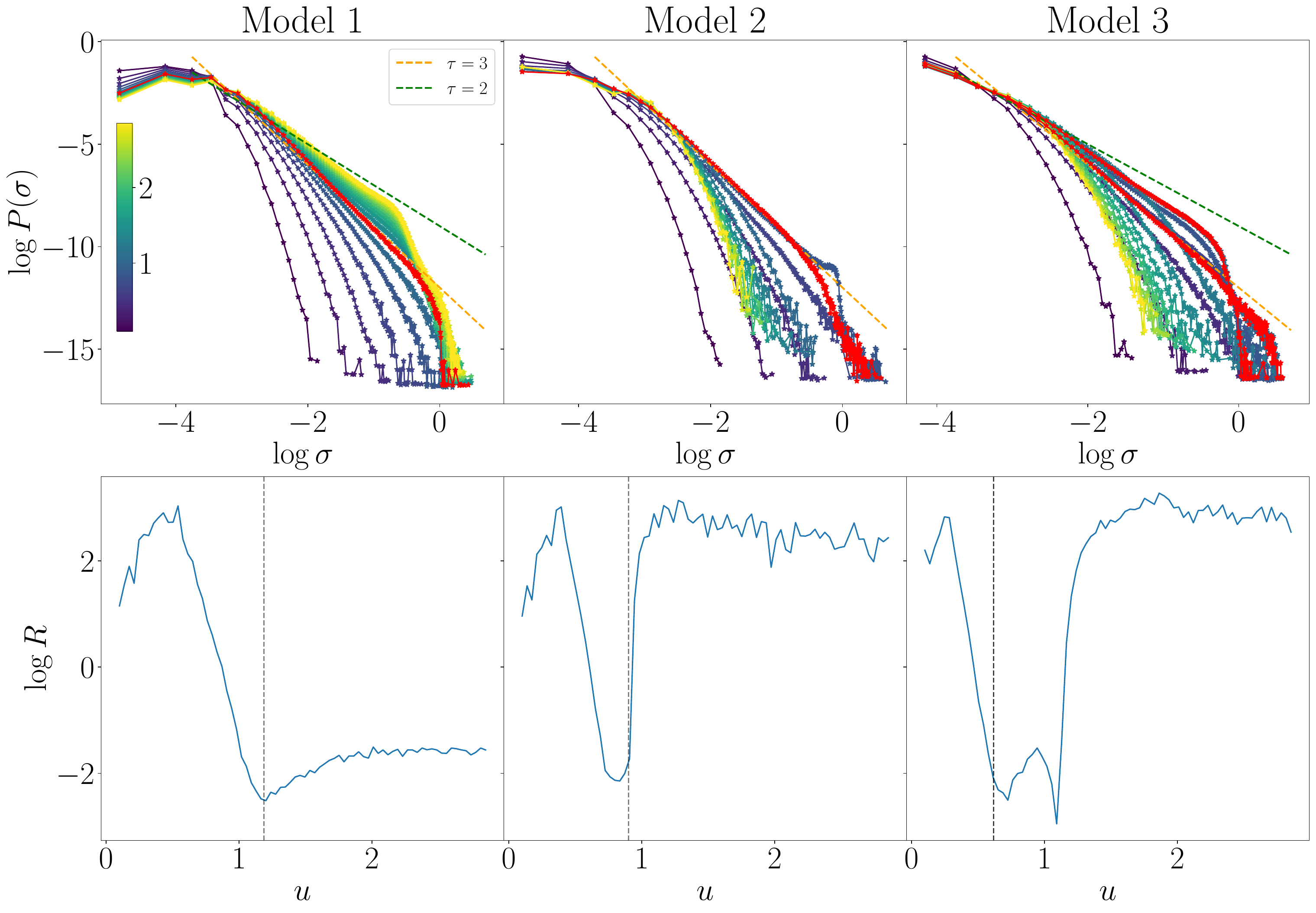}
  \caption{Top: We show the entropy impact probability distribution for the three models with phase transitions, with $u$ indicated by the color of the line. Purple corresponds to $u = 0.2$ and yellow corresponds to $u = 5.0$, and the color is linearly interpolated in between. In the appropriate regime, described below and in the text, the impact distribution function is linear on the log-log scale, indicating power-law behavior. In particular, in the appropriate fluctuating volume law regions, the avalanche distribution has a power between -2 and -3, indicated by the green and orange dotted lines, respectively. Additionally, the $u$ corresponding to $u_c$ and $u_d$ are highlighted in red. Bottom: We show the (logarithmic) least-squares error corresponding to the best-fit to the form $P(\sigma) = A \sigma^\tau$. We see that the error is small in the upper phase of model 1 and the intermediate phase of model 3. In the area-law phase of all three models, the entanglement is finite, so the entropy impact will be local, resulting in a poor fit to power-law behavior. Likewise, in the frozen volume-law of models 2 and 3, the system is frozen into regions protected from measurements except for at the boundary, resulting in measurements having only local effect on the entanglement substrate, again resulting in a poor fit to power-law behavior. Interestingly, for model 2, despite transitioning directly from an area-law to a frozen volume-law, we see that for $u \lesssim u_{d}$, a power-law fit appears to have small error, indicating that the measurement impact becomes nonlocal only very near the transition point.}
  \label{fig:avalanches}
\end{figure*}

For each update rule considered, we examine the entanglement entropy as the parameter $u$ is varied. For $u < 1$, we fix $p_m = 1$ and set $p_u = u$, and for $u > 1$, we fix $p_u = 1$ and set $p_m = 1/u$.  We also compute the mutual information defined as 
\begin{align}
    I_{AB} = S_{A} + S_{B} - S_{A \cup B}
\end{align}
for disjoint subsets of qubits $A$ and $B$. The mutual information is a commonly-applied tool for diagnosing the existence of a phase transitions in monitored quantum dynamics; a sharp peak in the mutual information is typically associated with a measurement induced continuous phase transition~\cite{li2019}. Using the entanglement scaling and mutual information, we identify three distinct phases and four distinct groups of models, shown in Fig.~\ref{fig:qrpm_model}(c). Data generated using the representative update rules are displayed in Fig.~\ref{fig:rep_modes} and Fig.~\ref{fig:rep_crossings}. In addition, in the Appendix A, we show the entropy substrate profile results for every possible configuration of update rules. 

The first model applies unitaries to the I, U, and L blocks, and measurements to T, P, and W blocks. This entanglement substrate for various $u$ is shown in Fig.~\ref{fig:rep_modes}. It exhibits a continuous phase transition $u_c \approx 1.21$ from area-law to volume-law entanglement scaling. Near the critical points, the entanglement entropy has logarithmic scaling in both space and time; that is, at early time $t\ll L$, the half-system entanglement scales as ${S_{[0,L/2]} = \alpha_t \log t}$ where we take the log to be base $e$ if not otherwise specified, and, after equilibration, the entropy saturates and scales with subsystem size as ${S_{[0,L_A]} \sim \alpha_s \log x}$, where $x = \frac{L}{\pi}\sin(\pi L_A/L)$. This logarithmic scaling is accompanied by a sharp peak in the mutual information, further indicating the presence of a critical point. The ratio of the scaling coefficients $\alpha_s$ and $\alpha_t$ is a universal quantity and equal to the dynamic critical exponent, i.e. ${z = \alpha_s/\alpha_t}$. Furthermore, we expect that the critical point is characterized by correlation lengths $\xi_\perp$ and $\xi_\parallel$ in the spatial and temporal directions respectively. They both diverge close to criticality as
\begin{align}
    \xi_\perp\sim (u-u_{c})^{-\nu_\perp},\quad \xi_\parallel\sim (u-u_{c})^{-\nu_\parallel}.
\end{align}
We compute $\nu_\perp$ and $\nu_\parallel$ by identifying data collapses of the steady state entanglement entropy in the spatial direction
\begin{align}
    |S_{[0,L_A]}(u) - S_{[0,L_A]}(u_{c})| = (u - u_{c})^{\nu_\perp} L_A,
\end{align}
and of the early time entanglement growth
\begin{align}
    |S_{[0,L/2]}(t, u) - S_{[0,L/2]}(t, u_{c})| = (u - u_{c})^{\nu_\parallel} t,
\end{align}
for appropriately chosen $\nu_\perp$ and $\nu_\parallel$. The dynamic exponent $z$ can equivalently be calculated as $z = \nu_\parallel/\nu_\perp$; we find, constraining $\nu_\perp = \nu_\parallel \equiv \nu$, we get a good data collapse, indicating that $z = 1$. We show the space and time scaling of the entropy substrate as well as the data collapse for models 1 and 2 in Fig.~\ref{fig:critical_modes}, and find that  $\nu = 1.42 \pm 0.13$, which is close to the value of $\nu = 1.3$ observed in hybrid random Clifford circuits~\cite{li2021}. The numerical details are presented in Appendix B. We also find $\alpha_s = 0.55$ and $\alpha_t = 0.60$, leading to $z = 0.91$. While we believe that the true value is likely to be $z=1$ based on the good data collapse for $\nu_\parallel$ and $\nu_\perp$, the scaling coefficients differ from the common result of $\alpha_s = \alpha_t = 0.76$ for the critical point of the hybrid random Clifford circuit, which manifests an emergent two-dimensional conformal symmetry~\cite{li2021}. These scaling coefficients also differ from the corresponding logarithmic prefactor to the RPM critical point of $\alpha_s = \alpha_t = 0.13$ \cite{degier2004}.

The volume-law phase exhibits significant fluctuations from one sample to another owing to the randomness present in the quantum circuits. Specifically, it is observed that for $t \ll L$, fluctuations along the temporal direction follow a power-law scaling form represented by
\begin{align}
    \delta S_{[0,L/2]}(t)=\sqrt{\langle [S_{0,L/2}(t)]^2 \rangle-\langle S_{[0,L/2]}(t)\rangle^2}\sim t^\gamma
\end{align}
where $\gamma=1/3$ (See Fig.~\ref{fig:kpz}). These fluctuations are also observed in the volume-law phases of various random circuits and are indicative of the Kardar-Parisi-Zhang (KPZ) universality class, which characterizes a broad class of surface growth models~\cite{kardar1986, corwin2011,Li_2023,han2023,Nahum_2017,Zhou_2019}. Furthermore, these fluctuations are observed in the entanglement entropy of the steady states, expressed as
\begin{align}
    \delta S_{[0,L/2]}=\sqrt{\langle [S_{0,L/2}]^2 \rangle-\langle S_{[0,L/2]}\rangle^2}\sim L^\gamma.
    \label{eq:kpz_fluc}
\end{align}
These fluctuations further contribute to a subleading correction in the steady state entanglement entropy with
\begin{align}
    S_{[0, L/2]} = a_1 L + a_2 L^{\gamma}.
\end{align}
We confirm this numerically in Fig.~\ref{fig:kpz}.

The second model applies unitaries to the {I, T, P, and W blocks}, and measurements to the remaining U- and L-blocks. It exhibits a phase transition characterized by a dicontinuous jump in the half-system entanglement density at the critical point $u_{d}$, as shown in the second column of Fig.~\ref{fig:rep_crossings}. As the parameter $u$ is increased, the entanglement density $S_{[0,L/2]}/(L/2)$ undergoes rapid growth from 0 to 1 when $u = u_d$. This becomes more pronounced and tends towards a step function as the system size is enlarged. This jump is accompanied by a large peak in the mutual information at the transition point. We further note that the volume-law phase for these models is characterized by vanishing fluctuations except for near $L_A = L/2$, indicating that the volume-law phase is exceptionally stable; we show the fluctuations vanish beyond the transition point in Fig.~\ref{fig:rep_crossings}. 

The stability of this almost maximally entangled volume-law phase can be understood by noting that unitary gates are applied to the W block for these modes. A region of several contiguous W blocks, corresponding to a maximally entangled region of qubits, is fixed under the update rules of the model, as this region cannot be acted on by measurements except at the boundaries, and unitary gates cannot change the entanglement in this region without violating the solid-on-solid restrictions. Therefore, this region of W blocks becomes ``frozen" when $u$ is sufficiently large. We observe that these frozen regions condense at the boundaries of the system, resulting in a length $\ell$ which is effectively protected from the action of the circuit dynamics. This region is only active at its boundary, where unitary gates and measurements can adsorb or desorb sites sites from the protected region, respectively. The endpoint of the frozen region undergoes a biased random walk towards $\ell = L/2$, leading to diffusive fluctuations in the entanglement entropy at early times.  Once the endpoint reaches $L/2$, the state evolves into an almost maximally entangled state and the entanglement entropy exhibits only small fluctuations when $L_A$ is close to $L/2$. As we reduce $u$, the drift velocity of the random walk decreases, causing it to take longer to reach the maximally entangled state. Eventually, at $u_d$, the random walk is unbiased, leading to $\ell \sim \sqrt{t}$. This picture is illustrated and numerically confirmed in Fig.~\ref{fig:frozen_mode}. In the steady state, the staircase edges growing from the left and right sides randomly walk across the width of the substrate until meeting at $L/2$. On average, the staircase edges sit a distance $L/4$ from the edges of the substrate, and thus the system has half of maximal entanglement. This can be observed by seeing that the entanglement density reaches a value of 1/2 at $u_d$ in the top row of Fig.~\ref{fig:rep_crossings} for model 2.

The third model applies unitaries to the I, T, L, and W blocks, and measurements to the remaining U and P blocks. It exhibits both a continuous phase transition from an area-law entangled phase into an intermediate fluctuating volume-law phase at $u_{c}$. At the critical point, the entropy scales close to logarithmically with respect to time and space, as in the first group of modes, although there are deviations from the logarithmic fit not observed in the first group; these can be seen in the third row of Fig.~\ref{fig:critical_modes}. Nonetheless, we are able to identify the scaling coefficients and critical exponents near the critical point, and find that model 1 and model 3 are similar; constraining $z = 1$, we find ${\nu = 1.32 \pm 0.09}$. These results are shown in Fig.~\ref{fig:critical_modes}. The intermediate volume-law phase also exhibits KPZ fluctuations, as confirmed by the numerical analysis presented in Fig.~\ref{fig:kpz}. As the parameter $u$ is increased further, the system goes through a phase transition into the frozen volume-law phase described above at $u_{d}$. Both transitions are accompanied by peaks in the mutual information, although the peak associated with the second transition into the frozen volume-law phase is several orders of magnitude larger than the first peak, and this large difference makes the lower peak difficult to numerically resolve for small system sizes; we observe that the two peaks can be identified concretely when $L \sim 512$, as shown in Fig.~\ref{fig:rep_crossings}.

The final model applies unitaries to all blocks except T, is drawn to the frozen maximally entangled volume-law phase independent of the parameter $u$. These systems have no entanglement transition at finite $u$. For very small $u$, the equilibration time becomes very large, so we instead start the system from a state which is close to maximally entangled and observe that the system is unable to descramble the information and reach an area-law for any $u$.  This maximally entangled state is reached by operating a random $L$-qubit random Clifford gate on the state $\ket{0}^{\otimes L}$. We note that the first three models are all able to descramble initially maximally entangled states when $u$ is small enough to leave the volume law. 

Finally, to further connect our model to the classical RPM, we analyze an entanglement substrate analogue of RPM avalanches. In particular, we compute the entropy ``impact" of a measurement performed at time $t$, defined as

\begin{align}
    \sigma &= \frac{1}{L}\sum\limits_i \left[h_i^{(t+1)} - h_i^{(t)}\right] \\
        &= \frac{1}{L}\sum\limits_i \left[S_{[0, L_i]}^{(t+1)} - S_{[0, L_i]}^{(t)}\right].
\end{align}
Thus, if a measurement affects the global entropy substrate, $\sigma \rightarrow 1$, whereas if the measurement only locally affects the entropy substrate, $\sigma \rightarrow 0$. We show the results in Fig.~\ref{fig:avalanches}. We find that the volume-law phase of model 1 and the intermediate phase of model 3 exhibit good fits to power-law behavior $P(\sigma) \sim \sigma^{\tau(u)}$ as characterized by the root mean square error

\begin{align}
    R = \sqrt{\frac{1}{L}\sum\limits_{\sigma = 1}^L (y(\sigma) - \hat{y}(\sigma))^2}
\end{align}
where $y(\sigma) = \log P(\sigma)$ and $\hat{y}(\sigma) = \tau \sigma + C$. The exponent $\tau$ depends on $u$ and varies between $-3 < \tau(u) < -2$ for the regions where the fit is valid. A power-law distribution of this nature suggests that the local measurement has the capability to suppress the entanglement entropy within a substantial subsystem. Nevertheless, due to the fact that the ``avalanche exponent" $-\tau(u)>2$, the measurement, on average, is only able to locally reduce entanglement. This power law behavior with large avalanche exponent has similarly been observed in the volume-law phase of the hybrid random Clifford circuit~\cite{li2019} and in the self-organized critical regime of the classical RPM~\cite{alcaraz2006}. Conversely, in the area-law and frozen volume-law phases of each model, the entropy impact is strictly local and $P(\sigma)$ decays faster than polynomially, indicated by the downward curve in Fig.~\ref{fig:avalanches}. 

\section{Discussion and Conclusion}
In this work, we consider a $(1+1)d$ quantum hybrid circuit model consisting of random Clifford unitary gates and projective measurements. For each timestep, unitary gates or projective measurements are applied to random sites depending on the local configuration of the entanglement entropy substrate according to prescribed feedback rules. This feedback mechanism, inspired by the RPM, leads to a rich phase diagram, shown in Fig.~\ref{fig:qrpm_model}(c). We observe three distinct phases: the area-law phase, the volume-law phase with a subleading power law correction, and the maximally entangled volume-law phase characterized by vanishing fluctuations at late time. The presence of these phases are determined by the update rules. The phase transitions between the area-law phase and the volume-law phase with a correction term are continuous and have conformal invariance. Conversely, the phase transition to a maximally entangled frozen volume-law phase is characterized by a discontinuous jump in the entanglement entropy density in the steady states and appears to be driven by the onset of diffusive growth of the protected boundary regions. In the volume-law phase, the entanglement growth exhibits diffusive fluctuations at early times, which can be interpreted using a random walk picture for the frozen staircase region. The drift velocity decreases as the measurement rate increases and drops to zero at the transition point, leading to diffusive dynamics at the transition point.

Furthermore, we compare our quantum model to the classical RPM by analyzing the distribution of avalanches, or, in the case of our model, the impact of measurements. We find that the effect of measurement on the entanglement substrate is local in the area-law phase, but for the fluctuating volume-law phase, the measurements affect regions of the entropy substrate on all length scales, indicated by a power-law distribution. We find that the power-law exponent varies between -3 and -2, in agreement with the self-organized critical phase of the classical RPM \cite{alcaraz2006, degier2004}. 

Interestingly, the classical RPM model can also be treated as a special non-unitary Clifford circuit. This is because there is a one-to-one correspondence between each height configuration and a non-crossing dimer configuration. In the latter, each dimer can be treated as a pair of Majorana fermions. Therefore, the RPM model can be mapped to a special Majorana fermion dynamics~\cite{Nahum2020}, which can be further described in terms of Clifford dynamics, where the measurement types and locations depend on the height profile developed in the previous time step~\cite{Sang_2021}. Currently, there is significant research on non-unitary free fermion dynamics~\cite{Chen_2020,Alberton_2021,Turkeshi}. Based on this perspective, we can potentially develop a deeper understanding of the phase diagram of the RPM model, most of which has been obtained numerically. We leave this for future study.

\section{Acknowledgements}
We acknowledge computational support from the Boston College Andromeda cluster. This research is supported in part by the Google Research Scholar Program and is supported in part by the National Science Foundation under Grant No. DMR-2219735 (E. H. and X. C.).

\bibliography{bibliography}

\newpage
\section{Appendix A: Results for all modes}

\begin{figure}[b]
  \centering
  \includegraphics[width=\linewidth]{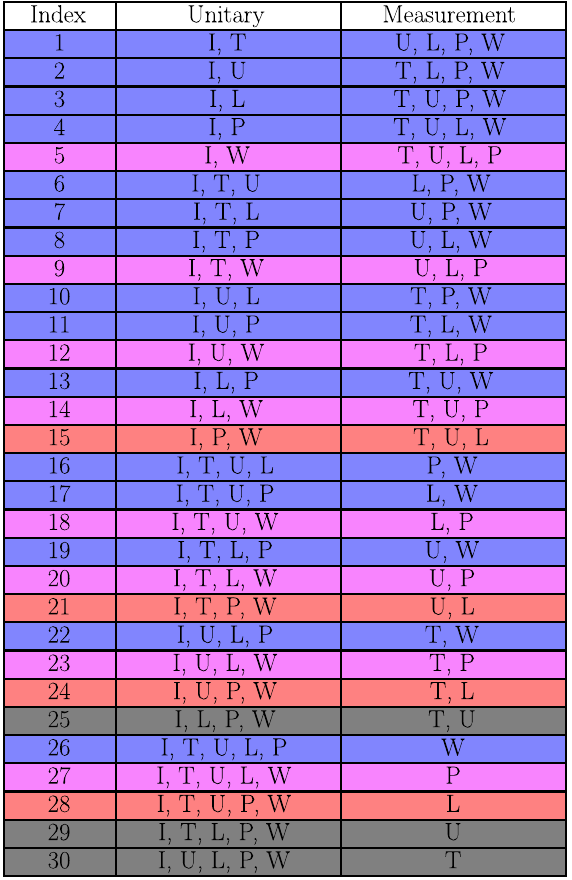}
  \caption{We enumerate each set of update rules possible for which either a unitary or measurement is applied to each block configuration in Fig.~\ref{fig:qrpm_model}(b). In the first column, we give the index of the update rules. In the second column, we list each block which a unitary is applied to. The remaining blocks are in the last column, which are the blocks for which a measurement is applied. The categorization of the models as discussed in the text is indicated by the color of the rows; blue, red, pink and grey correspond to the phase diagrams of models 1, 2, 3, and 4, respectively. }
  \label{fig:rule_table}
\end{figure}

\begin{figure*}
  \centering
  \includegraphics[width=\linewidth]{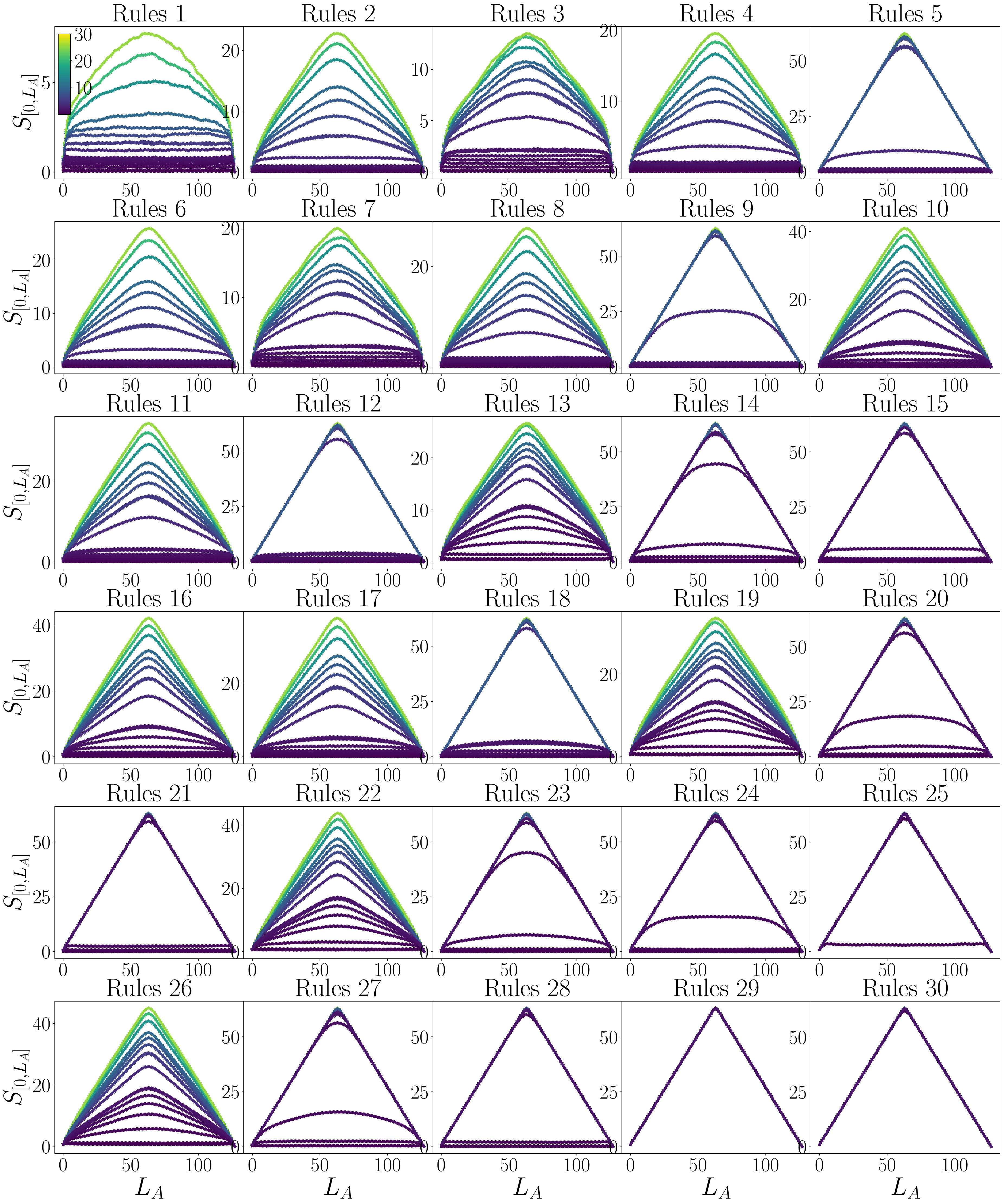}
  \caption{We show the entanglement scaling for all modes. For each mode, we choose $L = 128$ and evolve the system for $t = 3000L$ timesteps followed by $2000L$ timesteps of sampling. We average over 5 repetitions of this evolution. }
  \label{fig:all_modes}
\end{figure*}

In this section, we define a broader classification of update rules based on the the blocks in Fig.~\ref{fig:qrpm_model}(b) which have a unitary applied to them and which have a measurement applied. There are six distinct blocks, although starting from a flat substrate requires that a unitary be applied to the I block for any nontrivial entanglement growth to occur. This results in $2^5 = 32$ possible choices of blocks to apply unitaries to, of which we exclude the model which applies unitaries to every single block, and the model which applies unitaries to only the I block, resulting in 30 sets of update models. 

First, we enumerate every model in Fig.~\ref{fig:rule_table}. The models can be categorized into one of four phase diagrams, of which the models described in the text serve as representative members. Specifically, model 1 in the text has index 10 and has the same phase diagram as all models which does not apply unitaries to the W block.
Model 2 has index 21 and has the same phase diagram as all models which apply unitaries to the P and W blocks but not the L block, model 3 has index 20 and has the same phase diagram as all models which apply unitaries to the W block but not the P block, and model 4 has index 30 and has the same phase diagram as all models which apply unitaries to each of the L, P, and W blocks. This categorization is emphasized in the coloring of Fig.~\ref{fig:rule_table}

Finally, we show the entanglement substrate of each for each model for various values of $u$ in Fig.~\ref{fig:all_modes}.

\newpage
\section{Appendix B: Numerical techniques}
To compute the correlation length critical exponents for models 1 and 3, we employ similar techniques used by Ref.~\cite{skinner2019} to identify the $u_c$ and $\nu$ which lead to spatial and temporal data collapse. For the spatial (temporal) collapse, i.e. the data displayed in Fig.~\ref{fig:critical_modes}(d), we denote $y_L = |S_{[0,L/2}(u) - S_{[0,L/2]}(u_c)|$, and $x = (u - u_c)^\nu L$. We sample $y_L(x)$ for a finite set of $u_i$ corresponding to $x_{i,L} \in [x_{L,\text{min}}, x_{L,\text{max}}]$, and then calculate $y_L(x)$ as a continuous function with support $[x_{L,\text{min}}, x_{L,\text{max}}]$ by linear interpolation between sampled points. We then compute $\bar{y}(x) = \langle y_L(x) \rangle$, where the average is taken over the system sizes with support at $x$. Finally, we define the cost function

\begin{align}
    R_s(u_c, \nu) = \sum\limits_{L,i} [y_L(x_{i,L}) - \bar{y}(x_{i,L})]^2
\end{align}

Analagously, we define the family of curves $y_t$ for the early-time data, fixing $L = 512$. These are the curves shown in Fig.~\ref{fig:critical_modes}(b). Following the same procedure as above, we define the temporal cost function

\begin{align}
    R_t(u_c, \nu) = \sum\limits_{t,i} [y_t(x_{i,t}) - \bar{y}(x_{i,t})]^2
\end{align}

The total cost function is then 

\begin{align}
    R(u_c, \nu) = R_s(u_c, \nu) + R_t(u_c, \nu)
\end{align}

Minimizing this function yields the optimal $u_c$ and $\nu$. We perform the optimization using the Nelder-Mead method with randomly chosen initial conditions, repeating 100 times. The reported values are the values which correspond to the smallest value of $R(u_c, \nu)$ identified, and the reported error is the standard deviation across optimization trials. 

\end{document}